\documentstyle[aps,epsf,preprint]{revtex}
\tightenlines
\begin{document}
\draft
\title{Once more on mixing of the \boldmath{$a_0(980)$} and \boldmath{$f_0(980)$} mesons}
\author{N.N. Achasov\thanks{achasov@math.nsc.ru} and A.V. Kiselev\thanks{kiselev@math.nsc.ru}}
\address{Laboratory of Theoretical Physics,
 Sobolev Institute for Mathematics,
  Novosibirsk, 630090, Russia}

  \date{\today}
 \maketitle
\begin{abstract}
We show that a recent proposal by Close and Kirk fails to describe
$a_0(980)$ and $f_0(980)$ mixing when physical masses and widths
are included.
\end{abstract}


The mixing of the $a_0(980)$ and $f_0(980)$ mesons was discovered
theoretically as the threshold phenomenon in Ref.
\cite{achasov-79}. In Ref. \cite{achasov-81} a number of
experiments were proposed. There is also some additional
information in Ref. \cite{achasov-84}. Recently the interest in
the $a_0(980)$ and $f_0(980)$ mixing was renewed
\cite{achasov-97,kerbikov,close-00,kudryavtsev,close-01,kondratyuk,joe}.

In Refs. \cite{achasov-79,achasov-81,achasov-84,achasov-97}  is
shown that the $a_0(980)$ and $f_0(980)$ mixing effects can be
rather essential if a production amplitude of the resonance with
isotopic spin $I=1$ (or $I=0$) is considerably more than a
production amplitude of the resonance with $I=0$ (or $I=1$).  For
example, if the module of a production amplitude of the resonance
with $I=1$ (or $I=0$) is three times large as the module of a
production amplitude of the resonance with $I=0$ (or $I=1$) the
mixing effect can reach 10\%-20\%.

So, the conclusion of Ref. \cite{close-01}, that in the $K\bar K$
molecule model of the $a_0(980)$ and $f_0(980)$ mesons the mixing
effect can be as great as one likes even for equal production
amplitudes of the resonances with $I=0$ and $I=1$, is unexpected.
Unfortunately, this conclusion is not correct as we show below.

In Ref. \cite{close-01} the authors consider the molecule states
\begin{eqnarray}
&&\left |f_0\right > = \cos\theta\left |K^+K^-\right > +
\sin\theta\left |K^0\bar K^0\right >, \nonumber \\
&&\left |a^0_0\right > = \sin\theta\left |K^+K^-\right > -
\cos\theta\left |K^0\bar K^0\right >,
\end{eqnarray}
i.e., mixing of the states with the isotopical spin  $I=1$ and 0
\begin{eqnarray}
&&\left |f_0\right > = \cos\vartheta\left |f_0(I=0)\right > +
\sin\vartheta\left |a^0_0(I=1)\right >, \nonumber \\
&&\left |a^0_0\right > = \cos\vartheta\left |a^0_0(I=1)\right > -
\sin\vartheta\left |f_0(I=0) \right >\nonumber \\
&& \mbox{and the inverse equations}\nonumber \\
&&\left |f_0(I=0)\right > = \cos\vartheta\left |f_0\right > -
\sin\vartheta\left |a^0_0\right >, \nonumber \\
&&\left |a^0_0(I=1)\right > = \cos\vartheta\left |a^0_0\right > +
\sin\vartheta\left |f_0 \right >,\nonumber \\
\end{eqnarray}
where $\left |a^0_0(I=1)\right > = \left(K^+K^- -K^0\bar K^0
\right )/\sqrt{2}$, $\left |f^0_0(I=0)\right > = \left(K^+K^-
+K^0\bar K^0 \right )/\sqrt{2}$, $\vartheta = \pi/4 - \theta$,
$\theta = 30^\circ$.

 Suggesting that $\phi\to K^+K^-\to\gamma a_0$ and
 $\phi\to K^+K^-\to\gamma f_0$, they find
\begin{eqnarray}
&&  BR(\phi\to K^+K^-\to\gamma f_0)/BR(\phi\to K^+K^-\to\gamma
a_0)=  \cos^2\theta /\sin^2\theta.
\end{eqnarray}

In the case that the resonances overlap each other, in our case,
it is the fallacy. The point is that mixing the states with the
isotopical spin $I=1$ and 0 leads to the advent of the forbidden
decays $f_0^0\to\pi\eta$ and $a^0_0\to\pi\pi$,
\begin{eqnarray}
&& g(a_0^0\to\pi\eta) = g\cos\vartheta,\qquad
g(f_0^0\to\pi\eta)= g\sin\vartheta,\nonumber\\
&& g(f_0\to\pi\pi) = h\cos\vartheta,\qquad g(a^0_0\to\pi\pi)=
-h\sin\vartheta,
\end{eqnarray}
which lead to interference between the $a_0(980)$ and $f_0(980)$
resonances in the $\pi\eta$ and $\pi\pi$ spectra mass, which
compensates  mixing influence on the $\phi\to \gamma a_0$ and
$\phi\to\gamma f_0$ amplitudes practically completely.

Indeed, the $\phi\to \gamma (a_0 + f_0)\to\gamma\pi\eta$ and
$\phi\to\gamma (f_0 + a_0)\to\gamma\pi\pi$ amplitudes are of the
forms
\begin{eqnarray}
\label{a0}
 && A\left [\phi\to \gamma (a_0 + f_0)\to\gamma\pi\eta\right ]\propto
\left ( \frac{\sin\theta\cos\vartheta}{D_{a_0}(m)} +
\frac{\cos\theta\sin\vartheta}{D_{f_0}(m)} \right )=\nonumber\\
&& \frac{1}{\sqrt{2}}\left ( \frac{1}{D_{a_0}(m)} +
\cos\theta\sin\vartheta\sqrt{2}\left [\frac{1}{D_{f_0}(m)}-
\frac{1}{D_{a_0}(m)}\right ] \right )=\nonumber\\
&& \frac{1}{\sqrt{2}}\frac{1}{D_{a_0}(m)}\left ( 1 +
\cos\theta\sin\vartheta\sqrt{2}\left [D_{a_0}(m)- D_{f_0}(m)\right
]/D_{f_0}(m)\right )
\end{eqnarray}
and
\begin{eqnarray}
\label{fa0}
 && A\left [\phi\to \gamma (f_0 + a_0)\to\gamma\pi\pi\right ]\propto
\left ( \frac{\cos\theta\cos\vartheta}{D_{f_0}(m)} -
\frac{\sin\theta\sin\vartheta}{D_{a_0}(m)} \right )=\nonumber\\
&& \frac{1}{\sqrt{2}}\left ( \frac{1}{D_{f_0}(m)} +
\sin\theta\sin\vartheta\sqrt{2}\left [\frac{1}{D_{f_0}(m)}-
\frac{1}{D_{a_0}(m)}\right ] \right )= \nonumber\\
&& \frac{1}{\sqrt{2}}\frac{1}{D_{f_0}(m)}\left ( 1 +
\sin\theta\sin\vartheta\sqrt{2}\left [D_{a_0}(m)- D_{f_0}(m)\right
]/D_{a_0}(m) \right ),
\end{eqnarray}
where $D_R(m) = m_R^2 - m^2 -im\Gamma_R(m)$ is the propagator of
the $R$ resonance, $R=a_0\ \mbox{or}\ f_0$, $m$ is the invariant
mass of the $\pi\eta$ or $\pi\pi$ systems.

 So, when $D_{f_0}(m)= D_{a_0}(m)$, mixing
effect is absent at all!

The model of Ref. \cite{close-01} assumes that $D_{f_0}(m)$ is
very close to $D_{a_0}(m)$. But let us consider some figures. The
natural estimation of the mass difference is $m_{a_0}-m_{f_0}
=\cos 2 \theta \left (2m_{K^0}- 2m_{K^+}\right )= 4$ MeV. As for
the difference of the widths, $\Gamma_{a_0}(m_{a_0})-
\Gamma_{f_0}(m_{f_0}) =\pm 10$ MeV is the conservative estimation.
So,
\begin{eqnarray}
\label{modules}
 \left|\, \cos\theta\sin\vartheta\sqrt{2}\left [D_{a_0}(m)- D_{f_0}(m)\right
]/D_{f_0}(m) \right| < 0.08,\nonumber\\
 \left|\, \sin\theta\sin\vartheta\sqrt{2}\left [D_{a_0}(m)- D_{f_0}(m)\right
]/D_{a_0}(m) \right| < 0.05.
\end{eqnarray}
When calculating the right sides in the equations of
(\ref{modules}), we use $m=m_{f_0}$, $\Gamma_{f_0}(m_{f_0})= 50$
MeV in  the denominator of the first equation and $m=m_{a_0}$,
$\Gamma_{a_0}(m_{a_0})= 50$ Mev in the denominator of the second
equation.

So, the considerable mixing effect on the branching ratios is out
of the question. Nevertheless, let us consider the effect in more
detail. The branching ratios are the integrals of the spectra
$S_{a_0}(m)$ and $S_{f_0}(m)$ over m \cite{achasov-89}:
\begin{eqnarray}
\label{spectruma0}
&& S_{a_0}(m) = dBR[\phi\to\gamma (a_0 + f_0)\to\gamma \pi\eta\,,\, m]/dm
=\nonumber\\[6pt]
&&\frac{2}{\pi}\frac{m^2\Gamma(\phi\to\gamma
a_0\,,\,m)\Gamma(a_0\to
\pi\eta\,,\,m)}{\Gamma_\phi|D_{a_0}(m)|^2}\left |\, 1 +
\cos\theta\sin\vartheta\sqrt{2}\left [D_{a_0}(m)- D_{f_0}(m)\right
]/D_{f_0}(m)\,\right |^2
\end{eqnarray}
and
\begin{eqnarray}
 \label{spectrumf0}
&& S_{f_0}(m) = dBR[\phi\to\gamma (f_0 + a_0)\to\gamma \pi\pi\,,\, m]/dm
=\nonumber\\[6pt]
&&\frac{2}{\pi}\frac{m^2\Gamma(\phi\to\gamma
f_0\,,\,m)\Gamma(f_0\to
\pi\pi\,,\,m)}{\Gamma_\phi|D_{f_0}(m)|^2}\left |\, 1 +
\sin\theta\sin\vartheta\sqrt{2}\left [D_{a_0}(m)- D_{f_0}(m)\right
]/D_{a_0}(m)\,\right |^2.
\end{eqnarray}
Removing the module squares in Eqs. (\ref{spectruma0}) and
(\ref{spectrumf0}) it is easy to verify that the mixing
corrections have the same sign in Eqs. (\ref{spectruma0}) and
(\ref{spectrumf0}), i.e., the additional considerable compensation
of the mixing effects takes place in $S_{a_0}(m)/S_{f_0}(m)$.
Using the estimations (\ref{modules}), one can see that the mixing
effect correction in $S_{a_0}(m)/S_{f_0}(m)$ is less than 0.1,
which is to say that the mixing effect correction in
$BR[\phi\to\gamma (a_0 + f_0)\to\gamma \pi\eta]/BR[\phi\to\gamma
(a_0 + f_0)\to\gamma \pi\pi]$ is undoubted less than 10\%, i.e.,
less than the correction due to the difference of the volumes of
the $\pi\eta$ and $\pi\pi$ phase spaces.

Note that there is no tragedy with the relation between branching
ratios of $a_0$ and $f_0$ production in the $\phi$ radiative
decays.

The early predictions \cite{achasov-89} are based on the one-loop
mechanism $\phi\to K^+K^-\to\gamma a_0\to\gamma\pi\eta$ and
$\phi\to K^+K^-\to\gamma f_0\to\gamma\pi\pi$ at $m_{a_0}= 980$
MeV, $m_{f_0}=975$ MeV and $g_{a_0K^+K^-} = g_{f_0K^+K^-}$, that
leads to $BR(\phi\to\gamma a_0\to\gamma\pi\eta)\approx
BR(\phi\to\gamma f_0\to\gamma\pi\pi)$ .

But it is shown in Ref. \cite{achasov-97a} that the relation
between branching ratios of $a_0$ and $f_0$ production in the
$\phi$ radiative decays depends essentially on a $a_0 - f_0$ mass
splitting at $g_{a_0K^+K^-} = g_{f_0K^+K^-}$. This strong mass
dependence is the result of gauge invariance, the (photon
energy)$^3$ law on the right slope of the resonance.

Both SND and CMD detectors use the one-loop model $\phi\to
K^+K^-\to\gamma a_0$ and $\phi\to K^+K^-\to\gamma f_0$ in the data
treatment.  SND gives $m_{a_0} = 994^{+33}_{-8}$ MeV
\cite{snda0}(2000), $m_{f_0}=0.9698\pm 0.0045$ MeV
\cite{sndf0}(2000), CMD gives $m_{f_0}=0.969\pm 0.005$ MeV
\cite{cmd}.

Regarding the coupling constants, SND
 gives  $g_{a_0K^+K^-}^2/4\pi=1.05\pm^{0.36}_{0.25}$
\cite{snda0}(2000) GeV$^2$ and  , $g_{f_0K^+
K^-}^2/4\pi=2.47\pm^{0.73}_{0.51}$ GeV$^2$ \cite{sndf0}(2000). CMD
gives , $g_{f_0 K^+ K^-}^2/4\pi=1.49\pm 0.36$ GeV$^2$ \cite{cmd}.

So, there is no drastic difference between $g_{f_0K^+K^-}$ and
$g_{a_0K^+K^-}$.

As for the KLOE data, a detailed analysis is not presented up to
now\ \cite{kloe}.

The Ref. \cite{close-01} analysis of the ratio of production rates
in the central region at high energy by Pomeron - Pomeron
collision = $PP$ (isoscalar)
\begin{eqnarray}
\label{pomeron}
 \sigma (PP\to a_0)/\sigma(PP\to f_0) =
\frac{1-\sin 2\theta}{1+\sin 2\theta}= \frac{\sin
^2\vartheta}{\cos ^2\vartheta}
\end{eqnarray}
is not correct also.

As well as in the above example, proper allowance must be made for
interference between the $a_0(980)$ and $f_0(980)$ resonances in
the $\pi\eta$ and $\pi\pi$  mass spectra.

The $PP\to a_0 + f_0\to\pi\eta$ and $PP\to f_0 + a_0\to\pi\pi$
amplitudes are of the forms
\begin{eqnarray}
\label{Pa0}
 && A\left [PP\to a_0 + f_0\to\pi\eta\right ]\propto
\left (- \frac{\sin\vartheta\cos\vartheta}{D_{a_0}(m)} +
\frac{\cos\vartheta\sin\vartheta}{D_{f_0}(m)} \right )=\nonumber\\
&& \sin\vartheta\cos\vartheta\, \frac{D_{a_0}(m)-
D_{f_0}(m)}{D_{a_0}(m)D_{f_0}(m)}
\end{eqnarray}
and
\begin{eqnarray}
\label{Pf0}
 && A\left [PP\to f_0 + a_0 \to\pi\pi\right ]\propto
\left ( \frac{\cos ^2\vartheta}{D_{f_0}(m)} +
\frac{\sin ^2\vartheta}{D_{a_0}(m)} \right )=\nonumber\\
&& \frac{1}{D_{f_0}(m)}\left ( 1 - \sin ^2\vartheta\left
[D_{a_0}(m)- D_{f_0}(m)\right ]/D_{a_0}(m) \right ).
\end{eqnarray}

Estimating as in the above example we get
\begin{eqnarray}
\label{realpomeron} \sigma (PP\to a_0)/\sigma(PP\to f_0) < 4\times
10^{-3}.
\end{eqnarray}

Note that the experimental value, found in Ref. \cite{close-00},
\begin{eqnarray}
\label{exp}
 \sigma (PP\to a_0)/\sigma(PP\to f_0) =
 (8 \pm 3)\times 10^{-2}
\end{eqnarray}
differs from the prediction of Ref. \cite{achasov-81}
 \footnote{Note that this prediction is due to the strong coupling of
 the $a_0(980)$ and $f_0(980)$ mesons with the
$K\bar K$ channel.}
\begin{eqnarray}
\label{81} && \sigma (PP\to f_0\to a_0\to \pi\eta)/\sigma(PP\to
f_0\to\pi\pi)\approx
 BR\left (f_0\to\pi\eta\right )/BR(f_0\to\pi\pi)=\nonumber\\[6pt]
&&\frac{2}{\pi} \int \limits_{m_\pi + m_\eta}^{m_{max}}\left
|\frac {mM_{f{_0}a_{0}}(m)}{D_{a_0}(m)D_{f_0}(m)-
m^2M^2_{f{_0}a_{0}}(m)}\right |^2m^2\Gamma(a_0\to
\pi\eta\,,\,m)dm/BR(f_0\to\pi\pi) =\nonumber\\[6pt]
&& (0.5 - 2)\times 10^{-2}/BR(f_0\to\pi\pi)\approx (1.25 -
1.43)(0.5 - 2)\times 10^{-2}
\end{eqnarray}
by 1.71 - 2.46 experimental errors.  It is appropriate at this
point to recall that the data on $PP\to f_0(980)\to\pi^0\pi^0$ is
indirect. They are obtained in fitting a very complicated spectrum
with the help of the $f_0(980)$, $f_0(1300)$, $f_0(1500)$
resonances and a coherent background \cite{barberis}.

Emphasize that in the case of overlapping resonances, the
parameter of mixing is $M_{RR^\prime}(m_R)/\Gamma_R(m_R)$ and not
$mM_{RR^\prime}(m)/[D_{f_0}(m)-D_{a_0}(m)]$, where
$mM_{RR^\prime}(m)$ is the nondiagonal element of the polarization
operator describing the $R-R^\prime$ transition. In our case, in
the case of the isospin breaking transition, the very conservative
estimation is $\left
|M_{f_0a_0}(m_{f_0})/\Gamma_{f_0}(m_{f_0})\right|\alt 0.1$.

Note also that in the Close and Kirk model $mM_{RR^\prime}(m)=
\sin\vartheta\cos\vartheta (D_{f_0}(m)-D_{a_0}(m))$.

We thank very much J. Schechter, G.N. Shestakov and S.F. Tuan for
discussions.

This work was supported in part by RFBR, Grant No 02-02-16061.


\begin{references}
\bibitem{achasov-79} N.N. Achasov, S.A. Devyanin and G.N. Shestakov,
Phys. Lett. {\bf 88B} (1979) 367.
\bibitem{achasov-81} N.N. Achasov, S.A. Devyanin and G.N. Shestakov,
Yad. Fiz. {\bf 33} (1981) 1337 [Sov. J. Nucl. Phys. {\bf 33}
(1981) 715].
\bibitem{achasov-84} N.N. Achasov, S.A. Devyanin and G.N. Shestakov,
Usp. Fiz. Nauk {\bf 142} (1984) 361 [Sov. Phys. Usp. {\bf 27}
(1984) 161].
\bibitem{achasov-97}
N.N. Achasov and G.N. Shestakov, Phys. Rev. D {\bf 56} (1997) 212.
\bibitem{kerbikov}
B. Kerbikov and F. Tabakin, Phys. Rev. C {\bf 62} (2000) 064601.
\bibitem{close-00}
F.E. Close and A. Kirk, Phys. Lett. B {\bf 489} (2000) 13.
\bibitem{kudryavtsev}
A.E. Kudryavtsev and V.E. Tarasov, Pisma Zh. Eksp. Teor. Fiz. {\bf
72} (2000) 58 [JETP Lett. {\bf 72} (2000) 410].
\bibitem{close-01}
F.E. Close and A. Kirk, Phys. Lett. B {\bf 515} (2001) 13.
\bibitem{kondratyuk}
V.Yu. Grishina, L.A. Kondratyuk, M. B\"uscher, W Gassing and H.
Str\"oher, Phys. Lett. B {\bf 521} (2001) 217.
\bibitem{joe}
D. Black, M Harada, and J. Schechter, hep-ph/0202069.
\bibitem{achasov-89}
N.N. Achasov and V.N. Ivanchenko, Nucl. Phys. B {\bf 315} (1989)
465 .
\bibitem{achasov-97a}
N.N. Achasov and V.V. Gubin, Phys. Rev. D {\bf 56}, 4084 (1997).
\bibitem{snda0}
M.N. Achasov et al., Phys. Lett. B {\bf 438}, 441 (1998); {\bf
479}, 53 (2000).
\bibitem{sndf0}
M.N. Achasov et al., Phys. Lett. B {\bf 440}, (1998) 442 ; {\bf
485} (2000) 349.
\bibitem{cmd}
R.R. Akhmetshin et al., Phys. Lett. B {\bf 462} (1999) 380.
\bibitem{kloe}
A. Aloisio, Contributed paper to Lepton 2001, Rome, July 23-28;
hep-ex/0107024.
\bibitem{barberis}
D. Barberis et al., Phys. Lett. B {\bf 453} (1999) 325.
\end{references}
\end{document}